\documentclass[9pt,a4paper]{article}
\usepackage{blindtext}
\usepackage[T1]{fontenc}
\usepackage[utf8]{inputenc}
\DeclareUnicodeCharacter{2212}{-}
\usepackage{amsmath}
\usepackage{graphicx}
\graphicspath{{images/}}
\usepackage{subfigure}
\usepackage{hyperref}
\usepackage{authblk}
\usepackage[normalem]{ulem}
\usepackage[stable]{footmisc}
\hypersetup{
    colorlinks=true,
    linkcolor=blue,
    filecolor=magenta,      
    urlcolor=cyan,
}
\usepackage{vmargin}
\setmarginsrb{2.5 cm}{1.5 cm}{2.5 cm}{1.5 cm}{0.5 cm}{0.5 cm}{0.5 cm}{0.5 cm}
\date{\today}
\usepackage{setspace}
\usepackage{color,soul}
\title{Transport of particles through RO membrane in steady state condition }
\author{Rahul Karmakar}
\affil[1]{Department of Physics of Complex Systems, S. N. Bose National Centre for Basic Sciences, Block-JD, Sector-III, Salt Lake Kolkata 700106, India. \\
rahul.physics2017@gmail.com}

\author[1]{J Chakrabarti}

\affil[1]{Department of Physics of Complex Systems, S. N. Bose National Centre for Basic Sciences, Block-JD, Sector-III, Salt Lake Kolkata 700106, India.\\
jaydeb@bose.res.in}%

\begin{document}
%%%%%%%%%%%%%%%%%%%%%%%%%%%%%%%%%%%%%%%%%%%%%%%%%%%%%%%%%%%%%%%%%%%%%%%%%%%%%%%%%%%%%%%

\maketitle
%%%%%%%%%%%%%%%%%%%%%%%%%%%%%%%%%%%%%%%%%%%%%%%%%%%%%%%%%%%%%%%%%%%%%%%%%%%%%%%%%%%%%%%%%

\begin{abstract}

Reverse Osmosis(RO) membranes are widespread nowadays for separating the solvent from a solution. RO membranes are made of polymer matrix. Experiments show changes in relative interaction of solvent and solute with RO membrane matrix lead to changes in solvent permeation, solute rejection and fouling. Here we study microscopically separation of binary mixture through RO polymeric membrane using Molecular dynamics simulation to rationalize the experimental observations in terms a simple model. We find better solvent permeation and solute rejection with increasing interaction between solute and membrane with increase in fouling of solute inside membrane. We also find that stronger solvent membrane interaction reduces fouling. However, this leads to poorer performance of RO in terms of solvent permeation and solute rejection.

\end{abstract}
\newpage
\section{Introduction}
\large

Reverse osmosis(RO) is a method where external pressure is applied to overcome the natural osmotic pressure and enable solvent flow from  higher to lower solute concentration. As a result, RO technique is one of the most advanced ways to extract solvents from mixtures, and widely used in industries like water treatment\cite{ridgway2017molecular,song2017water}, biofuels\cite{hajilary2019biofuel}, petrochemical\cite{benito2010study}, pharmaceuticals processes\cite{radjenovic2008rejection} and so on. RO is used even in households for purification of water. RO for separation of smaller molecular size solute from organic solvents has also been developed in recent times\cite{kushida2022organic}. Such a widespread small to large scale utility of RO demands thorough understanding of RO membrane performance from microscopic considerations.

Thin Film Composites (TFC), consisting of polyamide layer  around $10-100$ nm thickness are widely used as RO membrane \cite{lau2012recent}. TFC membrane is prepared through interfacial polymerization(IP) process where two monomers in different phase(one in aqueous and other in organic) reacts only at the interface. Permeation or solvent recovery ($P$) in RO is estimated by how much solvent is transported from solution on one side of membrane to other side in presence of applied pressure overcoming osmotic pressure. The other important aspect is to reject solute (R) out of the solution allowing only solvent to permeate through the membrane.  Another important criterion of membrane is fouling ($F$). It is the measure of amount of solute accumulated inside the membrane during permeation. This degrades the future performance of membrane. The performance of membrane is to increase solvent permeation and solute rejection, while reducing the membrane fouling. 

Current studies on RO revolve around better design of RO membrane\cite{lim2021seawater}. The performance of membrane is affected by thickness, pore size distribution, roughness, hydrophilicity, chemical properties\cite{lau2012recent} etc. 
Experiments show that improving hydrophilic tendency by introducing different chemicals improve water permeation and salt rejection\cite{wang2010polyamide}. Studies also suggest that tuning TFC membrane using nanoparticles\cite{jeong2007interfacial}
as additives for better permeability. Incorporating ions like copper $Cu^{++}$ \cite{mehta2019poly} in membrane matrix show better pesticide removal from water and also better antibacterial properties. Performance in permeation increases introducing polypyrole layer in TFC membrane\cite{mehta2021polypyrrole}. Nonpolar mixtures are also separated using hydrophobic rich TFC membrane\cite{kushida2022organic,chau2018reverse}. Selective separation in both organic and polar solvents is recently developed using trianglamine macro-cycle as a monomer for cross-linked membranes through inter-facial polymerization(IP) process\cite{huang2020molecularly}.
Different permeation of solvent is observed in different solutions\cite{ding2015structure}. Experimental studies show anti-fouling properties by incorporation graphene oxide (GO) sheets in the polyamide layer\cite{inurria2019polyamide}. Theoretical studies have been carried out using full atomistic simulation using non-equilibrium molecular dynamics(NEMD)\cite{ding2015structure,shen2016dynamics}. Free energy profiles of water and ions are also studied inside thin films using the metadynamics technique to estimate barrier for permeation and rejection process\cite{gao2015understanding}.  However, there is yet no attempt addressing the generic features  of solute and solvent interaction with the membrane which control their transport through the RO membrane. 
 
\begin{figure}[!htb]
	\centering
	\includegraphics[width=10.0cm,height=5.3cm]{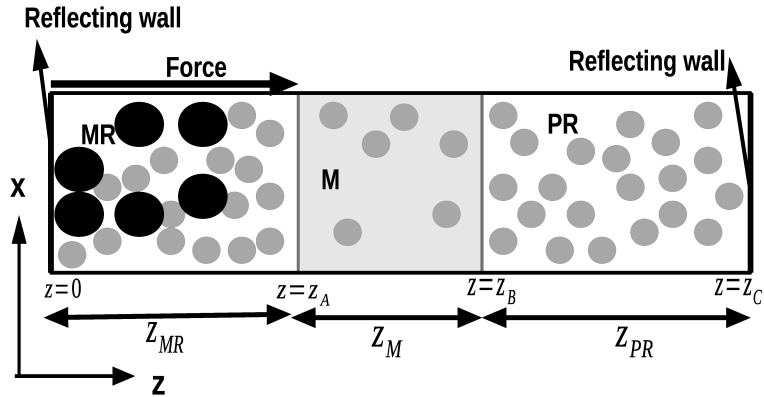}
	
	\caption{ Schematic diagram: M is the RO membrane in the middle solvated with S(grey) particles. MR is on left mixture reservoir containing S(grey) and L(black) particles. PR is on the right pure reservoir containing S particles only initially. Two reflecting wall is attached in two extreme end. Force is applied to MR.}

	\label{graph41 }
\end{figure}

Here, we investigate a simple microscopic model to study the solvent and solute motion in the steady state through a RO membrane in presence of an external force to overcome osmotic pressure. A schematic of the model is shown in Fig. \ref{graph41 }. We place the membrane in contact with two reservoirs: one of the reservoirs (Mixture reservoir, MR)($z=0$ to $z=z_{A}$) consists of binary mixture of LJ particles of different sizes, and  the other reservoir  only the smaller LJ particles (Pure reservoir, PR)($z=z_{B}$ to $z=z_{C}$).  The volume fraction of solvent in MR reservoir is lower than the PR reservoir, creating an osmotic pressure which is counterbalanced by an external force,  at $z = 0$  and falls linearly to zero at $z = z_{A}$. 

We study the system using molecular dynamics simulation. 
We fix size ratio of solute and solvent, driving force, and ambient temperature. There are several interaction energy parameters in the system: polymer bead-bead interaction, polymer bead-wall interactions, solvent-solvent interaction, solute-solute interaction, solvent-solute interaction which are also held fixed. We tune the relative strength of the interactions between the solvent and solute particles with the membrane. We find that the stronger interaction between solute and the membrane helps in better solvent permeation, solute rejection but with substantial membrane fouling. On the other hand, stronger solvent membrane interaction helps to decrease fouling. However, the solvent permeation and solute rejection also decrease concomitantly. 

\section{Model potential and simulation details} 

The non-bonded interaction between two monomers i and j of the polymeric strand with separation $r_{ij}$ is taken as the LJ 12-6 potential: 
\[
    V_{\alpha ,\beta}(r_{ij})= 
\begin{cases}\label{eq:41}
    4\epsilon_{\alpha,\beta} [ (\frac{\sigma}{r_{ij}})^{12} - (\frac{\sigma}{r_{ij}})^{6}],& \text{if } r_{ij} < 2.5\sigma\\
    0,              & \text{otherwise}
\end{cases}
\]Here $\alpha(=h,p)$ and $\beta(=h,p)$ stand for the bead types and $r_{ij}$ is separation between two beads. $\epsilon_{h,h}$ is the potential depth for interaction between h-h beads and $\epsilon_{p,p}$ that for the interaction between p-p beads. Here the interaction range parameters are $\sigma_{p,p}$ and $\sigma_{h,h}$.  The cross-interactions $\epsilon_{p,h}$ and $\sigma_{p,h}$ are taken to follow the Berthelot rule. 
The bonded interaction corresponding to stretching between two neighbouring beads at separation $r_{ij}$:
\begin{equation}\label{eq:42}
V_{bond} (r_{ij}) = \frac{1}{2} k_{b} (r_{ij} - r_{0})^{2}
\end{equation}
where $r_{0}=1.5\sigma_{h,h}$ is the equilibrium distance between monomers and $k_{b}$ the force constant. The change in bond angle also costs energy given by:
\begin{equation}\label{eq:43}
V_{angle} (\theta) = \frac{1}{2} k_{\theta} (\theta - \theta_{0})^{2}
\end{equation}
where $k_{\theta}$ the force constant and $\theta=\cos^{-1}( \frac{\vec{r}_{ij}\cdot{\vec{r}_{jk}}}{|\vec{r}_{ij}||\vec{r}_{jk}|})$ is the angle produced by three consecutive monomers i,j,k and $\theta_{0}$ is equilibrium angle, set to 114 degrees \cite{Opaskar1965TetrahedralAI}. 

Next, we specify the wall interactions with the polymeric beads. The $\beta (=wh,wp)$-th type wall interacts with a bead of $\alpha$-th type via the LJ 9-3 potential: 
\begin{equation}\label{eq:44}
V_{\alpha,w_\beta}(z_{i}) = \epsilon_{\alpha,w_\beta}[\frac{2}{15}(\frac{\sigma}{z_{i}})^{9} - (\frac{\sigma}{z_{i}})^{3}]
\end{equation}
Here  $z_{i}$ is the z-coordinate of i-th particle. The $\sigma$ values for all the interactions are taken to be the same but $\epsilon_{h,wh}>\epsilon_{p,wh} $ and $\epsilon_{p,wp}>\epsilon_{p,wh} $.  
 
The solvent S particles interact with each other via the LJ 12-6 potential with strength $\epsilon_{S,S}$ and length parameter $\sigma_{S,S}$. The solute L particles interact with each other with LJ 12-6 parameters $\epsilon_{L,L}$ and $\sigma_{L,L} $. Here the cross interaction parameters $\epsilon_{S,L}$ and $\sigma_{S,L}$ are taken to follow the Berthelot rule. $\epsilon_{S,h}$ and $\sigma_{S,h}$ and $\epsilon_{S,p}$ and $\sigma_{S,p}$ interaction parameters of S with h and p  polymeric beads respectively. Similarly $\epsilon_{L,h}$ and $\sigma_{L,h}$ and  $\epsilon_{L,p}$ and $\sigma_{L,p}$ are parameters for interactions of L with h and p beads respectively. 

We take $\epsilon_{h,h}$ as unit of energy. The unit of length is $\sigma_{S,S}=0.5$ nm. The unit of mass is the mass of water of density 1 $gm/cm^{3}$ in a sphere of diameter 0.5 $n m$. The polymeric bead diameter $\sigma_{p,p}=\sigma_{h,h}=2\sigma_{S,S}$. The solute particle diameter is taken $\sigma_{L,L}=1.8\sigma_{S,S}$. For all the wall interaction the length parameters, $\sigma=\sigma_{S,S}$. The fixed energy parameters are given in Table \ref{TR1}. The unit mass is
taken to be that of S particles $m_{S}$. The masses of all different particles are given in Table \ref{TR2}. Using the values of mass $m_{S}$, $\sigma_{S,S}$ and $\epsilon_{h,h}$,  we estimate a time scale, $\tau (=\sqrt{\frac{m_{S} \sigma_{S,S}^{2}}{\epsilon_{h,h}}})\sim 6$ picoseconds. Here we vary:  $\bar\epsilon_{S}=\epsilon_{S,h}/\epsilon_{S,p}$ and $\bar \epsilon_{L}=\epsilon_{L,h}/\epsilon_{L,p}$ at reduced temperature $T^{*} = \frac{k_{B}T}{\epsilon_{h,h}}=1$.

\begin{table}[h!]
\centering
\caption{Table for $\epsilon$}
\begin{tabular}{ |p{2.0cm}||p{2.0cm}||p{2.0cm}| }
 %\hline
 %\multicolumn{4}{|c|}{Country List} \\
 \hline
   Nature of interaction  & dimensionless variable &dimensionless value\\
 \hline
 h,h  & $\epsilon_{h,h}$ & 1.0   \\
 \hline
 p,p  & $\epsilon_{p,p}$ & 0.33   \\
 \hline
 h,wh  & $\epsilon_{h,wh}$ & 1.0   \\
  \hline
 p,wh  & $\epsilon_{p,wh}$ & 0.033   \\
  \hline
 p,wp  & $\epsilon_{p,wp}$ & 0.33   \\
  \hline
 h,wp  & $\epsilon_{h,wp}$ & 0.033   \\ 
  \hline
S,S & $\epsilon_{S,S}$ & 1.0 \\
  \hline
 L,L & $\epsilon_{L,L}$ & 1.0 \\

  \hline
\end{tabular}
%\caption{Table of chemical potential for different $\sigma_{tr}$ and temperature $T^{*}$.}
\label{TR1}
\end{table}

\begin{table}[h!]
\centering
\caption{Mass of particles}
\begin{tabular}{ |p{2.0cm}||p{2.0cm}| }
 %\hline
 %\multicolumn{4}{|c|}{Country List} \\
 \hline
   Mass of different particles  &dimensionless value\\
\hline
$m_{S}$  & 1.0   \\
\hline
$m_{L}$  & 5.832   \\
\hline
$m_{h}$  & 8.0   \\
 \hline
 $m_{p}$  & 8.0   \\

  \hline
\end{tabular}

\label{TR2}
\end{table}

The model involves several steps: (1) The model RO membrane(M) is prepared using 9 identical polymeric strands having randomly distributed 50 monomer beads of two different types in a number ratio 50:50 with overall packing fraction $\frac{\pi}{6}\frac{N}{V}\sigma_{h,h}^{3}=0.26$. One type of beads(h-beads) mutually interact more strongly compared to the other class of beads (p-bead) to mimic non-polar and polar moieties respectively. The network is confined between two asymmetric  walls, each of which is having favourable interaction with a given type of bead. This mimics the asymmetric interface between organic and aqueous phases where the membrane is stabilized in experimental conditions. We consider a rigid network with $k_{b}=k_{\theta}=k=100$ for simplification. We choose the interaction parameters such that the polymeric network is uniformly spread throughout the gap between two walls. (2) We solvate the membrane with the solvent (S) particles keeping the membrane in contact with two reservoirs of S particles where the confining walls are removed. (3) Then one of the reservoirs, (MR) is taken as a mixture of solutes (L) and solvents (S), while the other one, (PR) with pure S particles to study the reverse osmosis. The z width of membrane $Z_{M}=12$. The z width of MR and PR are taken $Z_{MR}=Z_{PR}=20$. 

We perform all the simulations using the LAMMPS package \cite{LAMMPS} and the Nose-Hover thermostat. The time step for integration is taken $0.001 \tau$. All the quantities are averaged over three different independent trajectories each  $1000000 \tau$ long. We take $L_{x}=30 $ and $L_{y}=20 $, in x- and y- directions respectively with the periodic boundary conditions (PBC) and no PBC in z-direction. The primary quantity we calculate is the packing fraction for different species of particles, $\eta_{\beta}^{\alpha}(t)=\frac{\pi}{6}\rho_{\beta}^{\alpha}(t)\sigma_{\beta, \beta}^{3}$, here $\rho_{\beta}^{\alpha}(t)$ is number density of particles of the $\beta$-th (= S,L) species at time $t$ in the $\alpha-th$ (=MR, M and PR) region. We quantify the membrane performance by the solute rejection rate, $R=100(1-\frac{\eta_{L,steady}^{PR}}{\eta_{L}^{MR}(0)})$; membrane fouling $F=100\eta_{L,steady}^{M}$; and solvent permeation or recovery $P = 100( \frac{\eta_{S,steady}^{PR}-\eta_{S}^{PR}(0)}{\eta_{S}^{PR}(0)})$. Here $\eta_{S}^{PR}(0)$ is initial partial packing fraction of S in PR and $\eta_{S,steady}^{PR}$ partial packing fraction of S in PR in steady state. $\eta_{L}^{MR}(0)$ initial partial packing fraction of L in MR and $\eta_{L,steady}^{PR}$ partial packing fraction of L in PR in steady state. $\eta_{L,steady}^{M}$ packing fraction of L inside M in steady state.

\section{Results}

\begin{figure}[!htb]
	\centering
	\includegraphics[width=5.3cm,height=10cm]{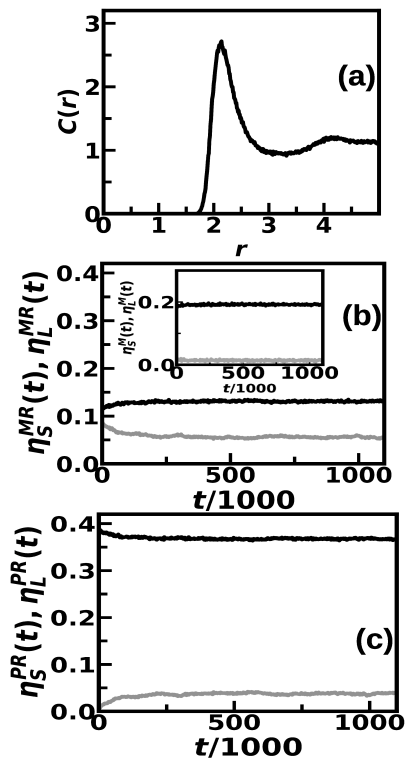}
	
	\caption{(a) Plot of inter-strand correlation function of beads
belonging to two different polymeric strands $C(r)$ versus r. (b) Packing fraction $\eta_{S}^{MR}(t)$(black line), $\eta_{L}^{MR}(t)$(grey line) as functions of time t with external force $F_{0}=1$. The parameters are $\bar{\epsilon}_{S}=1.74$, $\bar{\epsilon}_{L}=3.48$. Inset: Packing fraction $\eta_{S}^{M}(t)$(black line), $\eta_{L}^{M}(t)$(grey line) over time. (c) Plot for $\eta_{S}^{PR}(t)$(black line) and $\eta_{L}^{PR}(t)$(grey line) versus $t$ for the same parameters as panel (b).}

	\label{graph401 }
\end{figure}

We characterize interpenetrating network of identical polymeric strands by the inter-strand correlation function $C(r)$ defined in our previous works \cite{karmakar2022model} where $r$ is the cross-strand bead-bead distance.  The C(r) data in Fig.
\ref{graph401 }(a) shows a sharp first peak at $r=2$, equal to the polymeric bead diameter. This confirms that the monomers belonging to different polymeric chains stay close to each other which is a signature of an interpenetrating network. The membrane M is solvated by solvent S particles by placing it in contact with two identical reservoirs of S particles removing the confining walls. Further, we keep the z-coordinates of membrane beads fixed and the external force is set to zero.

The solvated membrane M is placed between a mixture reservoir (MR)  consisting of a binary mixture of 400 L particles ($\eta_{L}^{MR}(0) = 0.1)$ and 6000 S particles ($\eta_{S}^{MR}(0) = 0.26)$)and a pure reservoir PR  with 6400 S particles only($\eta_{S}^{PR}(0) = 0.28)$. Solvent and solute interaction parameters with the polymer beads ($\bar{\epsilon}_{S}=1.74$ with $\epsilon_{S,h}=1$ and $\epsilon_{S,p}=0.574$; and $\bar{\epsilon}_{L}$=3.48 with $\epsilon_{L,h}=2$ and $\epsilon_{L,p}=0.574$). We apply force to overcome osmotic pressure, $F(z)= F_{0}(1-z/z_{A})$. We use $F_{0}=1$, which corresponds to experimental range of pressure. We show packing fraction data $\eta_{S}(t)$ and $\eta_{L}(t)$ with time in different regions MR, M and PR. We observe in Fig. \ref{graph401 }(b) that the $\eta_S^{MR}(t)$ decreases in MR with time from $\eta_{S}^{MR}(0)$ till they saturate in steady state. We also observe $\eta_L^{MR}(t)$ decreases with t and then saturates in the steady state. In inset of Fig. \ref{graph401 }(b) $\eta_S^{M}(t),\eta_L^{M}(t)$ both saturates at large time. In Fig. \ref{graph401 }(c) both $\eta_S^{PR}(t),\eta_L^{PR}(t)$ increase and saturate in steady state. At large time in steady state, we measure packing fraction of S at PR, $\eta_{S,steady}^{PR}$, packing fraction of L at PR $\eta_{L,steady}^{PR}$ and packing fraction of L inside M $\eta_{L,steady}^{M}$. We observe that $\eta_{S,steady}^{PR}=0.36$, $\eta_{L,steady}^{PR}=0.037$, $\eta_{L,steady}^{M}=0.02$.  Using the steady state values we find for this case $R=62.46\%$, $P=32\%$, $F=2\%$. 
%\textcolor{red}{table not required; combine Fig.1 and 2}

\iffalse
\begin{figure}[!htb]
	\centering
	\includegraphics[width=10.3cm,height=10cm]{graph-40.eps}
	
	\caption{(a) Packing fraction $\eta_{S}(t)^{LR}$(black line), $\eta_{L}^{LR}(t)$(grey line) versus $t$ with external force $F_{0}=1$. The parameters are $\bar{\epsilon}_{S}=1.74$, $\bar{\epsilon}_{L}=3.48$. (b) Packing fraction $\eta_{S}^{M}(t)$(black line), $\eta_{L}^{M}(t)$(grey line) versus $t$. (c) Plots of  $\eta_{S}(t)^{RR}$(black line), $\eta_{L}^{RR}(t)$(grey line) versus $t$. The other parameters are same as panel (a).}

	\label{graph40 }
\end{figure}
\fi

\iffalse
\begin{table}[h!]
\centering
\caption{Steady state packing fraction of S,L in different region}
\begin{tabular}{ |p{2.0cm}||p{3.0cm}| }
 %\hline
 %\multicolumn{4}{|c|}{Country List} \\
 \hline
   Reservoir  & Steady state packing fraction\\
\hline
LR  & $\eta_{S,steady}^{LR}=0.13$, $\eta_{L,steady}^{LR}=0.05$    \\
\hline
M  & $\eta_{S,steady}^{M}=0.19$, $\eta_{L,steady}^{M}=0.02$  \\
\hline
RR  & $\eta_{S,steady}^{RR}=0.36$, $\eta_{L,steady}^{RR}=0.05$   \\

  \hline
\end{tabular}
%\caption{Table of chemical potential for different $\sigma_{tr}$ and temperature $T^{*}$.}
\label{TR4}
\end{table}
\fi

\begin{figure}[!htb]
	\centering
	\includegraphics[width=10.3cm,height=10.0cm]{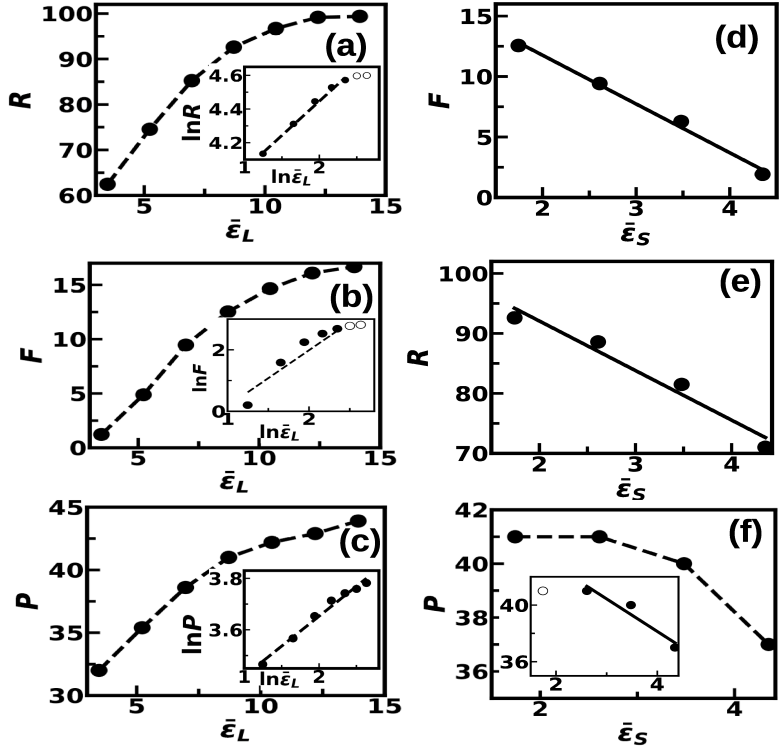}
	
	\caption{$R$ vs $\bar{\epsilon}_{L}$ plot with $\bar{\epsilon}_{S}=1.74$. (a) linear plot and inset, log-log plot.   $F$ vs $\bar{\epsilon}_{L}$ plot, (b) linear plot and inset, log-log plot.  $P$ vs $\bar{\epsilon}_{L}$ plot  (c) linear plot and inset, log-log plot. The parameters in (b) and (c) are the same as in (a). $\bar{\epsilon}_{L}=8.71$ for (d) to (f). (d) $F$ vs $\bar{\epsilon}_{S}$ and (e)$R$ vs $\bar{\epsilon}_{S}$ plots with the best fitted lines. $P$ vs $\bar{\epsilon}_{S}$ (f) linear and log-log plots in inset. The lines in the main panel is guide to the eyes unless otherwise stated explicitly. Inset: The line is the best fitted line. }

	\label{graph42 }
\end{figure}

We consider the effect of the interaction of the solvent and solute particles with the polymer beads over the RO performance. We study the case where both solute and solvent particles favour the h beads compared to the p beads, but they have different interaction strengths to the h-beads. Since the interactions with h-bead is stronger for both,  we have the restriction $\bar{\epsilon}_{L}>1$ and $\bar{\epsilon}_{S}>1$. Here we intend to stop L particles from moving from MR to PR. We study to this end the cases where $\bar{\epsilon}_{L}$ is varied and $\bar{\epsilon}_{S}$ (=1.74) is fixed. We observe that $R$ increases with increasing $\bar{\epsilon}_{L}$ in Fig. \ref{graph42 }(a) with, $R \sim \bar{\epsilon}_{L}^{0.4}$ dependence(log-log plot in inset of Fig. \ref{graph42 }(a)) and finally comes to saturation for large $\bar{\epsilon}_{L}$. The physical reason behind the observation is the following: $\bar{\epsilon}_{L}$ gets large, L particles find favourable to stay inside M. This scenario also suggests that $F$ increases as we observe from Fig. \ref{graph42 }(b) that $F$ also increases rapidly with $\bar{\epsilon}_{L}$, $F \sim \bar{\epsilon}_{L}^{1.82}$ (see log-log plot in the inset). We observe in Fig. \ref{graph42 }(c) that $P$ increases as $P \sim \bar{\epsilon}_{L}^{0.23}$(see the log-log plot in inset) with $\bar{\epsilon}_{L}$. Since L particles are stuck in MR and M, the S particle movement is facilitated from MR to PR. Thus, stronger solute and h-bead interaction, $\epsilon_{L,h}>\epsilon_{L,p}$ is favorable for membrane performance, but leading to fast growth in fouling.

We also consider increasing solvent-h bead interaction, where we increase $\bar{\epsilon}_{S}$, fixing $\bar{\epsilon}_{L}$ large (=8.71). In Fig. \ref{graph42 }(d) we plot $F$ with increasing $\bar{\epsilon}_{S}$. We observe $F$ sharply decreases linearly close to zero with increasing $\bar{\epsilon}_{S}$. However, Fig. \ref{graph42 }(e) shows that $R$ decreases linearly $\bar{\epsilon}_{S}$. On the other hand,  $P$ in  Fig. \ref{graph42 }(f)  decreases linearly as well with increasing $\bar{\epsilon}_{S}$ beyond $\bar{\epsilon}_{S}>2$. As $\bar{\epsilon}_{S}$ increases S particles are more favourable to stay inside the M. So L particles do not find favourable to stay inside M and also in MR due to forcing. They move towards PR which leads to a decrease in $R$, resulting in mixed phase in PR also. Thus, the increase in $\epsilon_{S,h}$ relative to $\epsilon_{S,p}$ impedes the performance of RO, although the  fouling can be reduced to almost zero. We do not show data larger value for $\bar{\epsilon}_{S}$ as we do not reach steady state in our calculation.

\begin{figure}[!htb]
	\centering
	\includegraphics[width=10.3cm,height=10.3cm]{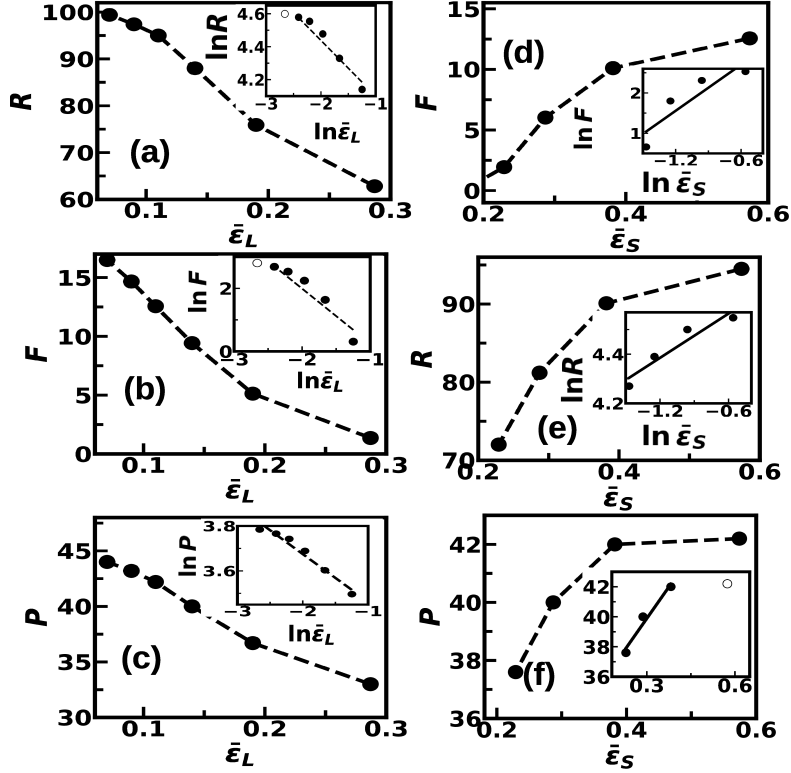}
	
	\caption{ (a)$R$ vs $\bar{\epsilon}_{L}$ plot with $\bar{\epsilon}_{S}=0.57$. linear plot and inset, log-log plot.   $F$ vs $\bar{\epsilon}_{L}$ plot, (b) linear plot and inset, log-log plot.  $P$ vs $\bar{\epsilon}_{L}$ plot, (c) linear plot and inset, log-log plot. The parameters in (b) and (c) are the same as in (a). $\bar{\epsilon}_{L}=0.114$ for (d) to (f). $F$ vs $\bar{\epsilon}_{S}$ plot, (d) linear and inset, log-log plots. $ R$ vs $\bar{\epsilon}_{S}$ plots, (e) linear and inset, log-log.  $P$ vs $\bar{\epsilon}_{S}$ (f) linear and log-log plots in inset. The line in the main panels is guide to the eyes. Inset: The line is the best fitted line. }

	\label{graph44 }
\end{figure}

Next, we consider the case where solvent and solute particles favour p-beads. In such cases both $\bar{\epsilon}_{S}$ and $\bar{\epsilon}_{L} < 1$. Here we fix $\bar{\epsilon}_{S}(=0.57)$ and vary $\bar{\epsilon}_{L}$. We observe in Fig. \ref{graph44 }(a) $R$ decreases with increasing $\bar{\epsilon}_{L}$ with $R \sim \bar{\epsilon}_{L}^{-0.33}$ (inset). Simultaneously, $F$ decreases rapidly with increasing $\bar{\epsilon}_{L}$, $F \sim \bar{\epsilon}_{L}^{-1.74}$ (Fig. \ref{graph44 }(b)). In Fig. \ref{graph44 }(c) we observe that $P$ also decreases with increasing $\bar{\epsilon}_{L}$ $P \sim \bar{\epsilon}_{L}^{-0.33}$ (inset). Since lower $\bar{\epsilon}_{L}$ denotes a strong L and p bead interaction, data suggests that stronger L and p interaction leads to better $R$ and $P$ but inferior performance in fouling. We further vary $\bar{\epsilon}_{S}$ in the regime $\bar{\epsilon}_{S}<1$. We observe in Fig. \ref{graph44 }(d) that $F$ increases with increasing $\bar{\epsilon}_{S}$ with exponent 1.5(inset) and so do both $R$(Fig. \ref{graph44 }(e)) and $P$ (Fig. \ref{graph44 }(f)). $R \sim \bar{\epsilon}_{S}^{0.3}$(inset) and $P$ increases linearly till $\bar{\epsilon}_{S}<0.45$(inset). Less fouling observed for $\epsilon_{S,h}<\epsilon_{S,p}$. However, this does not provide good RO performance for $P$ and $R$.

 Note that the relevant parameters in the study involve only the energy depth parameter of LJ interaction. This parameter is sufficiently robust and can be defined for any inter-molecular interaction having a well-defined minimum. Hence, it may be interesting to point out the implication of our model for realistic systems. The solution consisting of S and L particles both preferring the h-beads can be thought of the case as model for a solution of non-polar components. Our  calculation shows that the interaction of the non-polar solute should have stronger interaction with the non-polar moieties of the polymer membrane so that the non-polar solvent can easily flow through the membrane from mixture to the pure solvent. On the other hand, solution consisting of S and L particles both preferring the p-beads is a model for polar solution. Our calculations show that polar solute interaction with  polar moieties of the membrane should be stronger than the solvent interaction for easy passage of solvent from mixture to pure solvent. Thus, stronger solute membrane interaction results in larger solvent permeation and solute rejection but more membrane fouling. Our results at least qualitatively explain different experimental observations on how tuning the membrane properties affects  RO performance\cite{mehta2019poly}.

\section{Conclusion}
To summarize, we study the motion of particles from a mixture of solute and solvent particles to a pure solvent phase through a RO membrane driven by force to overcome osmotic pressure. We observe that the solvent recovery and solute rejection increase with increasing relative interaction of solute and polymeric membrane. Simultaneously, however, fouling of the RO membrane also increases. Although stronger solvent and membrane interaction leads to less fouling, this leads to impede solvent recovery and solute rejection as well. These results may serve guide for designing RO membrane  for targeted separation applications.

\bibliographystyle{vancouver}
\bibliography{paper.bib}

 \end{document}